# Streaming Multicast Video over Software-Defined Networks

Kyoomars Alizadeh Noghani, M. Oğuz Sunay

Özyeğin University, Department of Computer Science and Electrical Engineering, 34794 Istanbul, Turkey
kyoomars.noghani@ozu.edu.tr, oguz.sunay@ozyegin.edu.tr

*Abstract*—Many of the video streaming applications in today's Internet involve the distribution of content from a CDN source to a large population of interested clients. However, widespread support of IP multicast is unavailable due to technical and economical reasons, leaving the floor to application layer multicast which introduces excessive delays for the clients and increased traffic load for the network. This paper is concerned with the introduction of an SDN-based framework that allows the network controller to not only deploy IP multicast between a source and subscribers, but also control, via a simple northbound interface, the distributed set of sources where multiple-description coded (MDC) video content is available. We observe that for medium to heavy network loads, relative to the state-of-the-art, the SDN-based streaming multicast video framework increases the PSNR of the received video significantly, from a level that is practically unwatchable to one that has good quality.

## I. INTRODUCTION

ThisIt is predicted that approximately 73% of all IP traffic will be video by 2017 [1], of which some 14% will be from Internet video to TVs. Not surprisingly, streaming of live content is increasingly more prevalent on the Internet replacing the traditional means of TV broadcasting. One well-known method to alleviate the traffic load due to streaming video is to use IP multicast, which has been in existence for a long time. However, in today's networks, IP multicast has remained largely undeployed due to concerns on security, reliability and scalability, not to mention the requirement to have all routers in the network support the related protocols and be appropriately configured [2]. For this reason, application-layer multicast (ALM) has found prominence in the Internet where transmission of the content to the subscriber group is managed at the application layer and IP-unicast is used in the network layer for delivery with multiple copies of the same data transmitted over common links, incurring heavy loads on the Internet traffic. Additionally, compared to the IP multicast, ALM incurs longer latencies. The fundamental reason behind the prevalence of ALM despite its shortcomings is its immediate deployability, adaptability and updatability [3].

The rapid emergence of Software-Defined Networking (SDN) with significant industry backing [4] provides the perfect opportunity to implement IP multicast without any of its problems. Indeed, it is possible to construct, and maintain the multicast tree between a source and all its subscribers using a control application running on the logically centralized SDN-controller that has a global network view. The programmable nature of SDN allows for immediate deployability, scalability, adaptability, and updatability - traits all previously associated with ALM and not IP multicast. In this paper, we present an IP multicast application running on the SDN controller that also keeps track of the subscription activities via a simple northbound interface and illustrate its performance advantage.

IP Multicast is an ideal approach to mitigate the traffic load generated by streaming video services. A more efficient delivery of the video packets reduces the congestion probability in the network, which in turn improves the performances of both the corresponding streaming video system and all other concurrently running services on the network. In this paper, we first present an IP multicast framework for SDN, where we give a detailed description of the streaming video application, and its interaction with the SDN controller. Then we investigate how an actual implementation of IP multicast improves the streaming video performance relative to ALM in terms of streaming video quality-of-experience (QoE) metrics.

We require the streaming video application to be designed to satisfy the following conditions:

1) Support for different types of QoE-level based subscriptions should be present.
2) Resilience to network congestions and packet losses should be provided.
3) The video coding and decoding complexities should be as low as possible.

To answer these requirements, we consider an architecture that has the following properties:

1) Multiple streaming video servers, distributed across the network are deployed.
2) H.264-based Multiple Description Coding (MDC) is employed for video coding [5],[6] so that the same video content is described by multiple descriptions. With MDC, reception of one such description is sufficient for standard-quality playback, but delivery of multiple descriptions and a simple combining procedure of these descriptions prior to playback result in an increase in the video quality.

In the system studied herein, we consider a streaming server with two descriptions, available at two distinct locations of the network. We consider two subscription types: Standard and Premium. While a standard user is to receive content from one of the servers, premium users need to receive both descriptions for enhanced service quality. The server selection and associated multicast tree construction are orchestrated by a streaming-video specific IP-multicast application running on the SDN controller. In the subsequent sections we first provide a brief literature survey on various multicast studies and video delivery frameworks for SDN. We then present the proposed MDC-based streaming video service, followed by the SDN architecture on which it will operate and discuss the necessary interaction between the video application and the SDN controller. Next, we present experimental performance results for the video application using the QoE parameters of PSNR, and the number of pauses. Conclusions are drawn in the last section.

## II. LITERATURE REVIEW

By decoupling the control and data planes, and realization of the network control in software, SDN has been proposed to enable more agile and cost-effective networks [7]. Capitalizing on the benefits that SDN brings, there have been a number of studies on SDN multicasting in the literature.

In [8], an innovative way of managing IP multicast in overlay networks is proposed where the authors propose using OpenFlow instead of Internet Group Management Protocol (IGMP). [9] proposes a scalable, network-layer, single-source, inter-domain multicast framework by making use of a Locator/ID Separation Protocol (LISP) router overlay. [10] considers an IP multicast-based forwarding system, optimized for fast recovery in case of path failures. For each multicast group, the controller calculates two different multicast trees spanning all switches of the network. If a switch fails, the controller switches from the currently used tree to the complementary tree. [11] proposes a clean-slate approach for multimedia multicasting, where routes between the source and all of the subscribers are computed a priori with the purpose of speeding up the processing of multicast events over SDN framework.

There have also been a number of studies on streaming video applications for SDN. [12] considers a scalable video coded streaming video system where the base layer packets are given priority for guaranteed delivery while enhancement layers are routed either as lossy-QoS or best-effort flows. [13] presents an SDN-enabled content-based routing framework where Youtube flows are identified via Deep Packet Inspection (DPI) and are always forwarded via least congested links.

## III. MULTICASTING OVER SDN

The streaming video multicast framework presented herein is composed of two distinct parts: i) Streaming Video Multicast Service, ii) SDN Controller and the associated control application running on it. This is depicted in Figure 1.

Here, the Streaming Video Multicast Service maintains the identities and locations of the active servers and the corresponding descriptions they are multicasting. The service also maintains the up-to-date list of subscribers that are allowed to receive the service. The Streaming Video Multicast

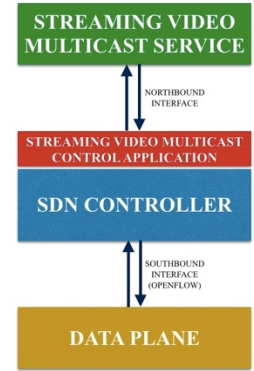

Fig. 1: Streaming Video Multicast Framework

Control Application running on the SDN controller is responsible from selecting the description(s) for each subscriber, establishing the corresponding route and maintaining the multicast tree for each description. The Streaming Video Multicast Service needs to ensure that the SDN controller has up-to-date information regarding both the subscriber and the video server identities. The SDN controller in return, needs to update the multicast service on whether a given subscriber has joined or left the multicast.

### A. Streaming Video Multicast Control Application

Provided that the identities of the subscribers and the servers are known at the controller, the control application needs to map clients with servers, compute routes and multicast trees for all clients and servers, respectively.

When constructing the multicast trees, two distinct optimization strategies could be considered when a subscriber wants to join the multicast group:

1) Minimize service impact on the network load
2) Maximize average streaming video quality

While the first strategy results in finding the server for which the addition of the subscriber to its multicast tree would result in the least number of additional branches in the tree, the second strategy finds the server that provides the highest QoE to the subscriber. In this paper we consider the second strategy.

We consider three different routing algorithms: Minimum Hop, Shortest Path (Dijkstra), and MinMax. Minimum Hop selects the server that is closest to the subscriber in terms of number of hops. In contrast, Shortest Path and MinMax select the paths that have the lowest sum link cost, and lowest maximum link cost from source to destination, respectively. Weights to the links are dynamic and are based on the traffic load they endure. In the proposed system, these weights are updated periodically to ensure good performance.

Once a server is selected for a subscriber and associated route is computed, the multicast control application adds this user to the corresponding multicast tree. This operation is repeated for every new subscriber. Subscribers may leave their multicast group politely or impolitely. When the leave is polite, the subscriber informs the multicast group a priori, but when it is impolite, the subscriber may leave with no notice.

In our framework, a user leaves its multicast group as a result of i) Crash/Shut down, ii) Disconnection from connected switch, iii) Service leave message.

To compute the link weights dynamically, the SDN controller periodically queries switches on their port statistics. The port statistics include the amount of received and/or transferred data. The controller then takes the average of $N=10$ previous statistics to determine a given link weight.

### B. Streaming Video Multicast Service

We refer to the streaming MDC-based video multicast servers as description providers (DPs). A newly launched DP sends a packet in multicast IP range containing information about its description which is always forwarded on to the controller. The multicast control application creates and stores a distinct multicast tree in the form of a data structure per DP. As soon as the source message arrives at the controller from a new DP, the control application establishes a new tree for that DP. Subsequently, the new DP is added to the list of available DPs so that for a new subscriber (or for an update for a current subscriber), when joint DP selection and routing is computed, this description is also considered.

At any given time, a DP may experience a failure. Either a proactive or a reactive solution may be developed for this scenario. For a proactive solution one of the following procedures may be implemented:

1) A back-up server may be made available for potential failures,
2) An alternate DP and associated multicast tree may be constructed for every subscriber a priori for fast tree switching.

For a reactive solution on the other hand, one of the following procedures may be implemented:

1) A new DP may be selected randomly after the failure is observed,
2) The best DP is computed for each client at the time of failure.

In this paper, we consider a reactive approach where the new best DP is selected upon the failure of the existing one.

### C. Subscribers

Subscribers join or leave the multicast streaming video service at any time via Join/Leave messages. When a new subscriber is to be added to the multicast tree, the control application conducts the following sequential procedure:

1) \It first checks whether the subscriber is already being served,
2) If not, it then checks whether the subscriber is to be served via communication with the multicast service,
3) Based on the routing algorithm in use, it selects the best DP for it,
4) It adds the subscriber to that DP's data structure,
5) It computes the necessary additional ports and/or branches to the multicast tree,
6) It pushes the corresponding forwarding rules to the switches using the OpenFlow protocol.

When a new subscriber joins a multicast tree, one of two scenarios may take place: i) joining the multicast tree may involve just the addition of a packet duplication rule to a single switch in the network, ii) joining the multicast tree might involve adding new switches and links to the multicast tree, in which case, rules for all affected switches are pushed.

Similarly, when a subscriber leaves the service, the control application conducts the following sequential procedure:

1) It first checks whether the subscriber is being served,
2) If so, it then removes the subscriber from its serving DP's data structure,
3) It then removes the port and/or switch and link from multicast tree,
4) It pushes the corresponding forwarding (expiration) rules to the switches using OpenFlow.

Similar to the subscriber join case, one of two scenarios may take place when a subscriber leaves the service: i) leaving the multicast tree may involve just the removal of the port of a switch from it if another multicast client is still attached to that switch, ii) leaving the multicast tree may involve removal of a switch and link from it. This case happens when the client, which is leaving the group, is a single leaf in multicast tree and multicast control application prunes the branch towards the leaf ancestor point, which is replicating packets on multiple ports.

In the proposed multicast service, it is possible for a subscriber to migrate from one DP to another. The main purpose for this migration is to increase user satisfaction from the service. Due to the dynamic nature of the network, it is possible that the DP which was the chosen as the best provider for a subscriber when it joined is no longer so. For this purpose, a separate thread periodically checks each client's best serving DP. If the current DP for one of the clients is no longer the best, the subscriber first leaves and then re-joins the service following the procedures outlined above.

## IV. ARCHITECTURE

The proposed streaming video multicast framework is built on three pillars:

1) IP multicast,
2) Multiple-Description Coding (MDC),
3) Software-Defined Networking (SDN).

MDC encodes the video into multiple, independently decodable streams where any description can be used to decode the media stream to provide error resilience to the system at the expense of a slight reduction in compression efficiency. Descriptions are distributed across the network to benefit from multipath routing. A similar benefit of error resiliency may be realized with Scalable-Video Coding with better coding efficiency. However, this improvement comes at the expense of the need for continuous careful orchestration of what different servers transmit and how packets from multiple servers are processed at the subscriber hardware, both of which require more advanced hardware realizations.

The use of IP multicast minimizes the unnecessary transmission of replicated packets in the network. Implementing multicast in network layer not only decreases the probability of network congestion but also increases the end-to-end packet delivery likelihood for all media and other services for clients at the same time.

In our implementation, the streaming video content is MDC-coded with 2 distinct descriptions. We have a number of servers in the network, each server streaming one of the two descriptions. Two classes of subscriptions are possible for the service: Standard and Premium. The premium users subscribe for a high quality streaming experience. They achieve this via reception of both descriptions. For this purpose, premium users need to be on two distinct multicast trees at a given time. When delivery of one description fails, the premium user will still be able to continue its playback, albeit, at a reduced quality level. The standard users on the other hand, subscribe for a basic quality streaming experience. At a given time, the standard user receives only one description for playback, and thus belongs to only one multicast tree. When the delivery from this tree fails, the standard user experiences a pause until the failure is corrected or the user is migrated to a new server. The premium user experiences a pause only when both trees experience failures. Figure 2 illustrates a comprehensive view of MDC streaming video using multicast over SDN.

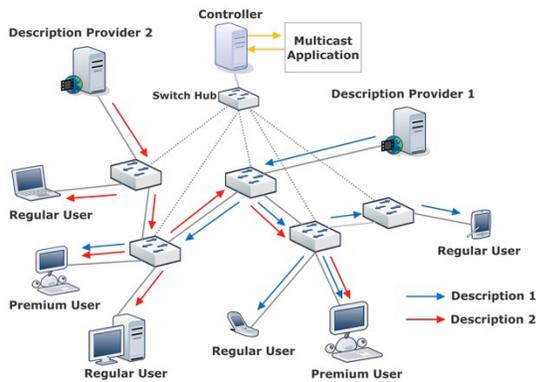

Fig. 2: General Perspective

## V. EXPERIMENTAL SETUP & EVALUATION

We conduct experiments to assess the performance of the proposed architecture when Minimum Hop, Shortest Path and MiniMax routing algorithms are deployed. To assess the benefit of SDN-based IP multicast MDC video streaming, we also investigate the performance of ALM for both SDN as well as non-SDN networks as benchmarks. In the SDN network, end-to-end routes are established also for ALM by the controller, which has global network view. The non-SDN network, on the other hand, is today's Internet, where routes are computed in a distributed manner by the individual switches, which have local network views of their neighbourhoods. The non-SDN network is emulated by enabling the Learning Switch module in the Floodlight Controller as this module invokes a behaviour akin to today's standard switches.

We conduct the experiment on Mininet version 2.0 in four different topologies with 15-20 switches, implemented with Open vSwitch 1.4. In the topologies, each switch is connected to an average of 2.67 other switches. We assume that each link has a bandwidth of 100 Mbps. The topology sizes are selected so that investigation of a heavily congested network is possible. Congesting bigger topologies with higher link bandwidths is difficult with limited memory. We use 10-320 cross traffic generators to create real traffic patterns to assess the system performance at different network loads.

For each network topology, the experiment is conducted for a period of 20 minutes. An additional 5 minutes is set aside to initialize the emulation testbed and 2 minutes between each emulation to calm down the CPU and memory usage. All experiments run over an IBM Server with 12 cores of CPU and 28 Gigabytes of RAM.

DPs stream packets with the exact size and rate of the actual streamed video which is 15fps with an average 1000 kb/s bit rate, but instead of transmitting the video data, we let the DPs populate the packets with parameters which help us analyze and track them further a posteriori. These parameters include information regarding the source DP, when packet is sent, packet sequence number, etc. Clients capture packets, parse the associated data, stamp the packet with reception time and save it. Also clients record the time they join and leave the streaming service. The QoE metrics are subsequently calculated for each subscriber based on the individual packet reception dynamics.

During the initialization phase, the cross-traffic generators start congesting the network for the first minute. The DPs start streaming the two emulated descriptions of the video for another minute after which the subscribers start joining the service. Artificial traffic between cross clients and servers are generated using 4 real patterns of HTTP, FTP, audio and video conference (Skype) and video streaming (YouTube), captured using Wireshark. The main purpose of the cross-traffic generation is to congest the network and since this is not achievable by TCP due to its inherent congestion control mechanism, all cross-traffic packets are transmitted using UDP in the experiment. Each cross traffic server transmits data to a predetermined receiver that remains status throughout the experiment. The servers randomly select one of the four above mentioned traffic patterns for a duration of 1024 packets, and then switch randomly to one of the other patterns, and so on.

In all experiments, the sequence of joining and leaving for both standard and premium users follow a predetermined pattern for fair comparison of the results. In the experiment, we assume that each client stays in the multicast group for 45 seconds after its Join message. Afterwards, the clients randomly choose to either submit a new request (Leave request if they are already being serviced and vice versa) or remain in their current state for another 45 seconds. We assume that the probability of submitting a leave request is 20% and the probability of submitting a join request is 80%. A higher probability is considered for subscribing to the service since the clients may capture more packets this way, which in turn improves the evaluation accuracy. In the experiment, we consider 10 multicast clients, which are distributed across the network. 50\% of the clients are assumed to be standard users, and 50\% are premium users. Finally we investigate the following QoE metrics:

1) Packet Loss
2) Pre-Roll Delay
3) PSNR

### D. Packet Loss

We first investigate the percentile loss of video packets due to congestion in the network. The results are depicted in Figure 3. We observe that as the cross-traffic in the network increases (loaded network), the ALM performance becomes significantly worse than SDN-based IP multicast performance. While SDN with ALM performs better than a non-SDN with ALM, both have significantly worse performances than the SDN with IP multicast. This result confirms that while SDN is essential in reaping the gains of the IP multicast architecture, it is not sufficient on its own. It acts as an enabler to easily implement architectures that would be difficult, or even impossible, in today's Internet, which in turn provides significant performance gains.

Of the three routing algorithms, MiniMax incurs the lowest packet loss. However, the performance difference between them is not very large. This result is dependent on the topologies on which the experiment is run. MiniMax may achieve more significant gains over different topologies where there are more available paths between switches.

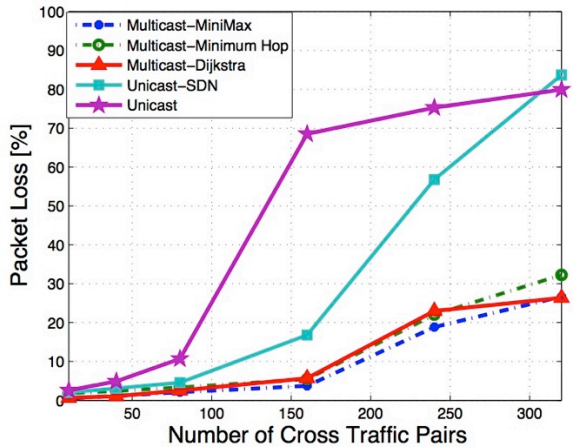

Fig 3.: Packet Loss

### E. Pre-Roll Delay

Pre-Roll Delay is defined as the difference between the time when a user subscribes for service and first packet for playback is received at the receiver. Table I tabulates the observed pre-roll delays for SDN-based IP multicast with different routing algorithms, as well as SDN and non-SDN-based ALM for different network congestion levels. The table lists the values for the top performing 90% and 100% subscribers. The table entries with dashes correspond to the case where there are less than 90% or 100% of the users that can receive the service in that scenario as a number of the users are denied service completely due to congestion. We observe that while all service options perform similarly for lightly loaded networks, IP-multicast continues to serve over 90% of its clients even when the network is very heavily congested, albeit, with increased average pre-roll delays.

### F. PSNR

Peak signal-to-noise ratio (PSNR), defined as the ratio between the maximum possible power of a signal and the power of corrupting noise that affects the fidelity of its representation, is commonly used to quantify the quality of a video. The well-known Foreman test video used for PSNR calculation. A representative video frame from Foreman is depicted in Figure 4.

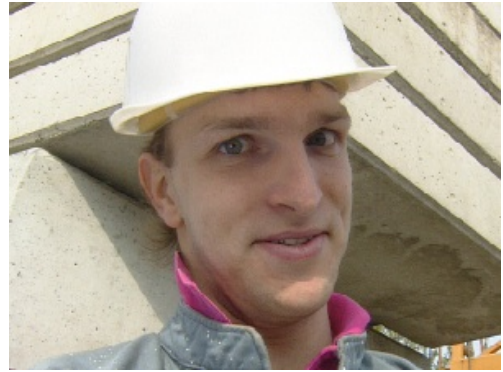

Fig 4.: A Video Frame from the Test Video: Foreman

Due to its large size, a video frame is usually segmented into multiple packets for transmission over the network by the DP. If a packet is not received, the video player at the client end invokes a simple error concealment procedure in which the missing part is replaced from the video frame preceding it. Then, the PSNR value for the video is calculated using the following sequential procedure:

1) Sort the packets which are received for each video frame,
2) Check if any packet is missing,
3) If so, invoke error concealment,
4) Combine packets to populate the video frames,
5) When all frames are generated at the receiver, compare them with their original counterparts.

The PSNR values are shown in Figure 5 for the test video. We observe that for a network with medium load, IP multicast - with either of the routing algorithms - provide near lossless video quality of 37 dB and 29 dB, for premium and standard users, respectively. The corresponding non-SDN ALM values on the other hand are very low, indicating a non-watchable video. The PSNR loss with IP multicast due to increased congestion remains tolerable throughout, but the ALM performance results in non-watchable videos throughout.

| Cross Traffic | Multicast-MiniMax | | Multicast-MinHop | | Multicast-Dijkstra | | Unicast-SDN | | Unicast | |
|---|---|---|---|---|---|---|---|---|---|---|
| | 90% | 100% | 90% | 100% | 90% | 100% | 90% | 100% | 90% | 100% |
| 10 | 0.3 | 0.3 | 0.3 | 0.3 | 0.3 | 0.3 | 0.3 | 0.3 | 0.3 | 0.3 |
| 40 | 0.3 | 0.3 | 0.3 | 0.3 | 0.3 | 0.3 | 0.3 | 0.3 | 0.3 | 0.3 |
| 80 | 0.3 | 0.4 | 0.3 | 0.3 | 0.3 | 0.3 | 0.3 | 0.3 | 0.5 | - |
| 160 | 0.5 | 0.7 | 0.4 | 0.6 | 0.4 | 3.4 | 0.5 | - | - | - |
| 240 | 20.1 | - | 30.1 | - | 23.8 | - | - | - | - | - |
| 320 | 28.7 | - | - | - | 35.7 | - | - | - | - | - |

TABLE I: Observed Pre-Roll Delay Values as a Function of Network Congestion

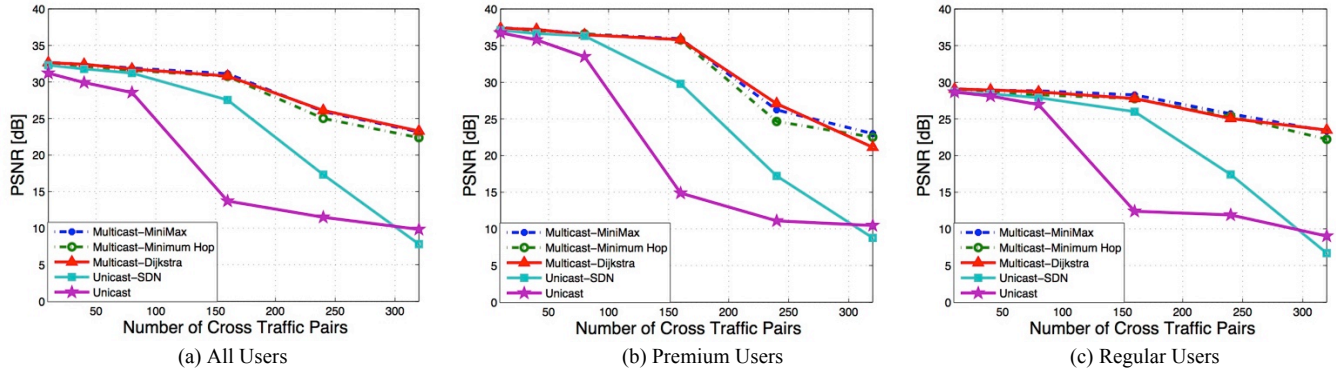

Fig. 5: Foreman – PSNR values for all clients

(a) All Users    (b) Premium Users    (c) Regular Users

## VI. CONCLUSION

Video has become one of the most prominent applications of the Internet. Many of the video streaming applications involve the distribution of content from a CDN source to a large population of interested clients. However, widespread support of IP multicast has been unavailable to a large extent due to technical and economical reasons, all stemming from the non- programmable nature of today's Internet. As a solution, streaming multicast video is commonly operated using application layer multicast. However, this technique introduces excessive delays for the clients and increased traffic load for the network. This paper introduces a multiple-description coded streaming video multicast framework that can be easily realized using software-defined networking. We observe that for medium to heavily loaded networks, relative to today's solution of application layer multicast in a non-SDN network, the SDN-based streaming multicast video framework increases the PSNR of the received video significantly; from a level that is practically unwatchable to one that has good quality. Unlike today's solution of ALM, over 90\% of the subscribers receive the video service, albeit at a higher pre-roll delay. We conclude that SDN is a powerful enabler of easily deployable, programmable, powerful network control, with which it is possible to observe significant performance gains.

## REFERENCES

[1] Cisco Systems Inc., "The zettabyte era – trends and analysis," *White Paper,* May 2013.

[2] C. Diot, B. Levine, B. Lyles, H. Kassem, and D. Balensiefen, "Deployment issues for the IP multicast service and architecture," *IEEE Network*, vol. 14, no. 1, pp. 78–88, Jan. 2000.

[3] M. Hosseini, D. Ahmed, S. Shirmohammadi, and N. Georganas, "A survey of application-layer multicast protocols," *IEEE J. Commun. Surveys and Tutorials*, vol. 9, no. 3, pp. 58–74, Mar. 2007.

[4] Open Networking Foundation, "Software-defined networking: The new norm for networks," *ONF White Paper*, Apr. 2012.

[5] M. Pereira, M. Antonini, and M. Barlaud, "Multiple description coding for Internet video streaming," *Proc. IEEE Image Process. Conf. (ICIP)*, vol. 3, Barcelona, pp. 281–284, Sept. 2003.

[6] R. Yang, S. Zheng, and T. Cao, "H.264 based multiple description video coding for Internet streaming," *Proc. Multimedia Technol. Conf. (ICMT)*, Ningbo, pp. 1–4, Oct. 2010.

[7] N. McKeown, T. Anderson, H. Balakrishnan, G. Parulkar, L. Peterson, J. Rexford, S. Shenker, and J. Turner, "OpenFlow: Enabling innovation in campus networks," *ACM SIGCOMM Comput. Commun. Rev.*, vol. 38, no. 2, pp. 69–74, Mar. 2008.

[8] Y. Nakagawa, K. Hyoudou, and T. Shimizu, "A management method of IP multicast in overlay networks using openflow," in *Proc. Hot Topics in Software Defined Networks (HOtSDN)*. Helsinki: ACM, 2012, pp. 91–96.

[9] F. Coras, J. Domingo-Pascual, F. Maino, D. Farinacci, and A. Cabellos- Aparicio, "LCAST: Software-defined inter-domain multicast," *Elsevier J. Comput. Networks*, vol. 59, pp. 153–170, Feb. 2013.

[10] D. Kotani, K. Suzuki, and H. Shimonishi, "A design and implementation of OpenFlow controller handling IP multicast with fast tree switching," in *Proc. IEEE Symp. Applicat. and the Internet (SAINT)*, Izmir, pp. 60–67, July 2012.

[11] L. Bondan, L. F. Müller, and M. Kist, "Multiflow: Multicast clean-slate with anticipated route calculation on OpenFlow programmable networks," *Elsevier J. Applied Computing Research*, vol. 2, no. 2, pp. 68–74, 2013.

[12] H. Egilmez, S. Civanlar, and A. Tekalp, "An optimization framework for QoS-Enabled adaptive video streaming over openflow networks," *IEEE Trans. Multimedia*, vol. 15, no. 3, pp. 710–715, Apr. 2013.

[13] M. Jarschel, F. Wamser, T. Hohn, T. Zinner, and P. Tran-Gia, "SDN-based application-aware networking on the example of youtube video streaming," in *Proc. European Workshop on Software Defined Networks (EWSDN)*, Berlin, pp. 87–92, Oct. 2013.